# A comparative study on communication structures of Chinese journals in the social sciences

Ping Zhou,[a,b] [*] Xinning Su,[c] and Loet Leydesdorff [d] [*]

[a] Centre for R&D Monitoring (ECOOM),
Katholieke Universiteit Leuven,
Leuven, Belgium
[b] Institute of Scientific and Technical Information of China
Beijing, China
[c] Chinese Social Sciences Research Evaluation Center, Nanjing University,
Nanjing, China
[d] Amsterdam School of Communications Research (ASCoR), University of Amsterdam, Amsterdam, The Netherlands.

**Abstract**

We argue that the communication structures in the Chinese social sciences have not yet been sufficiently reformed. Citation patterns among Chinese domestic journals in three subject areas—political science and marxism, library and information science, and economics—are compared with their counterparts internationally. Like their colleagues in the natural and life sciences, Chinese scholars in the social sciences provide fewer references to journal publications than their international counterparts; like their international colleagues, social scientists provide fewer references than natural sciences. The resulting citation networks, therefore, are sparse. Nevertheless, the citation structures clearly suggest that the Chinese social sciences are far less specialized in terms of disciplinary delineations than their international counterparts. Marxism studies are more established than political science in China. In terms of the impact of the Chinese political system on academic fields, disciplines closely related to the political system are less specialized than those weakly related. In the discussion section, we explore reasons that may cause the current stagnation and provide policy recommendations.

---

[*] These authors contributed equally to this work.

**Introduction**

The percentage of world share of publications with a Chinese address increased spectacularly in the natural and life sciences, but remained dramatically behind in the social sciences. Figure 1 provides the percentages for the *Science Citation Index* and the *Social Science Citation Index*, respectively. The line at the bottom shows the percentage of world share of China in the *Social Science Citation Index*, whereas the exponential curves show the Chinese percentages (with and without Hongkong). For the orientation of the reader, comparable figures for Germany are also provided (dashed lines).

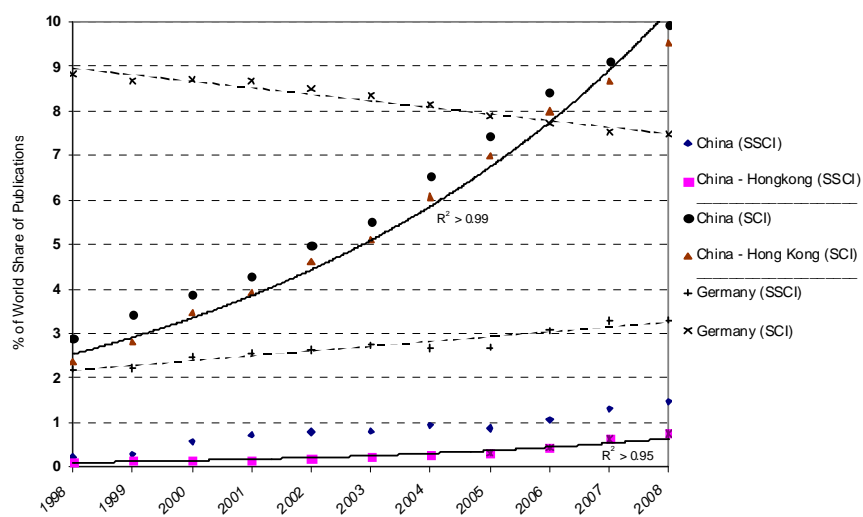

**Figure 1**: Percentages of world share of China (with and without Hong Kong) and Germany in the Science Citation Index and Social Science Citation Index, respectively. (Articles, reviews, proceedings papers, and letters, included; integer counting.)

Whether one includes Hong Kong or not, the percentages of world share in the natural and the life sciences shows exponential growth during the last ten years. In the *Social Science Citation Index*, however, the inclusion of Hong Kong makes a large difference. In recent years (since 2005), growth is due to an increasing share of publications with an address from Mainland China, but the huge difference with growth in the natural sciences remains to be explained. We added the figures for Germany (which also has a large national literature in the social sciences), for reference. In our opinion, the different nature and orientation of the social sciences in China, in particular, are based on communication structures that impede participation in the international arena (more than in Germany, for example).

What causes the huge differences in performance of China between these two intellectual domains? Is this an effect of using the international databases for the measurement? Do social scientists in China perhaps perform better domestically, than internationally?

**Social sciences in China**

The social sciences study human society and individual relationships in and to society. This orientation generates close links between the social sciences, and the ideological and political systems of specific countries and world regions. Starting to appear at the end of the nineteenth, and the beginning of the twentieth century in China (Xia & Zhang, 1999; Dong, 1999), the Chinese social sciences have two typical features. First, the disciplines originated in the West, and were imported into China. Therefore, localization of scholarly insights in the Chinese context has always been a significant task for Chinese scholars (Hicks, 2004). Second, marxism had a special position in the Chinese academic community and political system, since the foundation of the People's Republic in 1949. These two significant features led to characteristics in the Chinese social sciences different from those of the West.

Furthermore, the Chinese social sciences suffered from the political movements at the system level, more than the natural sciences. For example, the ten-year long Cultural Revolution (1966-1976), almost destroyed the foundations of the Chinese social sciences. While the country was extemely self-isolated during the Cultural Revolution, and intellectual resources were destroyed on a massive scale, international scientific relations did not disappear completely in the natural and life sciences. The reason for this is that these sciences can immunize themselves against political contexts, to a certain extent. The social sciences, however, by their very nature, participate in the discourses that they study, and are able to maintain only soft boundaries with their political environments. Furthermore, the aspiration of the Communist Party of leading the country, using marxism as a basis for social-scientific theorizing, set a framework for such sciences as economics and political science.

In 1978, the Chinese government adopted its new Opening-up and Reform Policies. Education and research in the social sciences were gradually reconstructed. Academic talents, trained both domestically and internationally, continuously enlarge China's research force in the social sciences (Xinhuanet, 2009; CPG, 2007a, 2007b). Where would these (large numbers of) scholars publish? In addition to the international publication arena, a majority of the Chinese scholars regularly publish in domestic journals.

In 2006, China published 2,339 journals in the social sciences and humanities (Jiang, 2007; Ren 2007). However, the *Social Science Citation Index* 2007 contained no Chinese journals at all, with only five journals registered in China, but published by organizations not from the mainland of China. In 2000, two Chinese databases similar to the *Social Science Citation Index* were created: the *Chinese Social Science Citation Index* (*CSSCI*) and the *Chinese Humanities and Social Science Citation Database* (*CHSSCD*). In 2007, the *CSSCI* covered 493 journals (*CSSCI*, 2009), whereas the *CHSSCD* covered 793 journals in 2001 (CNKI, 2009). These databases contain citation information for journals, and therefore enable us to study the communication patterns among Chinese scholars publishing in the domestic domain of the social sciences at the aggregated level.

A series of studies on citation relations among Chinese journals in the sciences (Leydesdorff & Jin, 2005; Ren & Rousseau, 2002, 2004; Moed, 2002; Zhou & Leydesdorff, 2007) show that communication among Chinese scholars in the sciences is less embedded in scholarly literature, in terms of citations and references, than communication among scholars in the international journals. In the Chinese scientific community, furthermore, knowledge seems to flow predominantly from high-quality international journals to the lower-level domestic ones. International journals have a higher rank in this hierarchy than their Chinese counterparts: knowledge flow and exchanges between Chinese and international scholars consequently are not balanced (Zhou, 2009).

Initially, we wished to explore in this study how Chinese scholars in the social sciences communicate with their international counterparts, by analyzing the citation environments of domestic Chinese journals, in relation to international communication as operational, in terms of the citation patterns in the *Social Science Citation Index*. However, upon exploration of the data, it became clear that this topic currently can not be operationalized in terms of the databases, because no single Chinese journal covered by the domestic *CSSCI* is also covered by the international *SSCI*. Although five journals in the *SSCI* are registered in China, these journals are managed by organizations from either Hong Kong, or the USA. The so-called non-source citations of Chinese authors in domestic journals to international literature are also very sparse.

In other words, the Chinese and international communication systems in the social sciences are almost completely uncoupled in terms of the coverage in the databases. Paradoxically, this complete isolation of the Chinese domestic and international communication circuits may make the results of a comparison between the two systems more interesting because one can assume the absence of interactions caused by cross-coverages between the *SSCI* and the *CSSCI*. Although Chinese scholars publish in international journals covered by the *SSCI*, their role is not significant enough to affect the international patterns of citations among these journals. China's world share of publications in the *SSCI* is still low. In 2006, the *SSCI* covered 1,176 Chinese publications, which is not even 1.5% of the world total (Zhou *et al.*, 2009). Of these, approximately 50% is still from Hong Kong. The contribution of international scholars to Chinese domestic journals in the social sciences is virtually absent.

Although we could not study the citation relations between the international and domestic domains—because there is not sufficient data available—it is possible to focus instead on what the different patterns of citation relations among domestic and international journals can teach us about the structure of communications at the disciplinary level. We compare the communication patterns among Chinese scholars publishing in the domestic domain with citation relations among journals included in the *SSCI*. We show that the journal structures have not sufficiently been reformed, remained largely in the old regime, and therefore stagnated. Some signs of recent improvements, however, can also be signalled, and policy recommendations will be formulated accordingly.

**Materials and methods**

For the purpose of this project, the data contained in the *CSSCI* was elaborated in a format comparable to the *Journal Citation Reports* (*JCR*) of the *SSCI,* so that we were able to use previously developed routines for analyzing the latter database. In 2007, the *CSSCI* covered 493 journals, which is around 20% of the whole set of Chinese journals in the humanities and social sciences. The *SSCI* covered 1,866 journals in the same year (*JCR Social Sciences Edition*, 2007).

Journal-journal citation relations are extracted from the corresponding databases by applying a set of dedicated routines, previously developed by Leydesdorff and Cozzens (1993). Using these routines, a citation environment can be obtained, based on a specific (seed) journal most relevant to the subject under study. Since a journal may refer to (or cite) and can be cited by other journals, one can analyze a citing and cited environment of a journal, respectively. This enables the analyst to distinguish citing patterns—which indicate the knowledge base of a set—from being-cited patterns, or impact environments of journals. Only journals surpassing 1% of the "total citing" or "total cited" counts of the seed journal will be used in the "citing" or "cited" environment of the seed journals, respectively. When a "citing" or "cited" environment contains too few (e.g., fewer than three) journals, using this threshold of 1%, we enlarge the inclusion by lowering the threshold to 0%, and thus including all citing or cited journals.

Citing/cited relations among journals contain valuable information. In addition to citation impact and the influence of "self-citations" within journals, one can also indicate whether a discipline or field is specialized using the reference (or citing) patterns of journals within a closed set representing a specific discipline. Questions naturally emerge: what is specialization of science and what causes specialization? Scientific specialization can be considered as a result of narrowing one's focus of study in a discipline, and thus subfields or specialties can emerge at the aggregated level (Leydesdorff, 2007).

Sociological and historical studies account for scientific specialization in a number of ways. Some scholars consider specialization as a combined consequence of social changes, changes in focus of pioneering studies, and conceptual or cognitive changes (Edge & Mulkay, 1976; Lemaine et al., 1976; Mulkay & Edge, 1976; Worboys, 1976). Ben-David and Collins ([1966] 1991) regard the creation of a new scientific specialty as a consequence of scientists carving out a new professional niche in an effort to create a new social role. Phase transition—preparadigmatic, paradigmatic, postparadigmatic (Weingart, 1997)—may lead to the closure of philosophical debates and re-specification of contexts of application (Etzkowitz & Leydesdorff, 2000; Gibbons *et al*., 1994; Leydesdorff, 2010).

Price ([1963] 1986) and Beaver & Rosen (1979a, 1979b) attribute scientific specialization to the accumulation of knowledge. As an increasing number of authors become involved in science, and more and more journals publish ever more papers, each new generation of scientists is confronted with a larger body of scientific literature. Price ([1963] 1986) estimated that the number of publications was doubling every fifteen years, and hypothesized that the optimal size for a scientific research community is between 100 and 200 publishing scientists.

Only by narrowing their area of research, and thus creating a new specialty, are scientists able to effectively manage the continuously growing literature. Beaver & Rosen (1979b) also indicated that as knowledge accumulates at an increasing rate, fields may begin to split into subfields. These developments, in turn, spawn new and more restrictive societies and journals (i.e., specialized journals). Closed clusters of journals within a discipline thus can be used as an indicator of specialization. When it comes to a citation network among journals, a discipline/field can be considered as well specialized if the reference patterns are focused across journals within the discipline/field or, in other words, self-referential to a disciplinary identity (Van den Besselaar & Leydesdorff, 1996).

In addition to studies on restructuring of knowledge within a discipline/field, relations between disciplines have also been investigated. For example, Cronin and Meho studied intellectual transfer in Information Studies (Cronin & Meho, 2008), by comparing intellectual flow between disciplines as a kind of trade. They used the "balance of trade" to measure the healthiness of a discipline/field, and stated that "a discipline that is a net exporter of ideas to others can be said to have a healthy balance of trade. … A strong discipline may be one that has a positive trade balance, but that need not necessarily be the case. Conversely, a discipline with a poor export record is not thereby a failing field".

Instead of studying the "balance of trade" between disciplines, the current paper focuses on specialization within a discipline, so as to address the following issues: 1) what is the overall difference between the Chinese domestic and international scholarly communication systems, and 2) do the political system and ideology still play a role in the Chinese scholarly communication system?

The first question will be investigated at the database level by comparing the *CSSCI* and the *SSCI,* using several indicators. In order to be able to investigate the second issue, we classified the journals into two groups, based on the expected intensity of the possible impact of national political systems or ideology on specific disciplines: strongly and weakly affected disciplines. Political science is, for example, selected to represent journals in this first group. The *CSSCI* further classifies journals focusing on marxism studies as "Marxism", under which are twelve journals. We selected journals from both "Political Science" and "Marxism" in the *CSSCI* subject categories to explore citation patterns of journals that may strongly be affected by national politics and ideology.

For disciplines in group two (i.e., weakly affected by China's political system), we set three basic conditions for selecting journals, given the limited coverage of the *CSSCI*. The first condition is that scholars in these selected fields must show international publication behaviour. Second, corresponding subject categories should exist in both the *CSSCI* and the *JCR* of the *SSCI*. Third, the *CSSCI* subject category must cover a sufficient number of journals so as to ensure reliability of the results.

The *CSSCI* has 25 subject categories with approximately 20 journals per subject category. However, the number of journals subsumed under some of these subject categories is low. For example, only seven journals are classified as psychology, and four journals as statistics. Hence, subject categories satisfying the above two conditions are limited.

**Table 1.** Number of journals covered by the *CSSCI* and *SSCI* in 2007 in the selected subject categories.

|  | *Political Science* | *Economics* | *Library & Information Science* |
|---|---|---|---|
| ***CSSC*** | 50 | 72 | 18 |
| ***SSCI*** | 93 | 191 | 56 |

After combining these conditions and data source limitations, we selected the subject categories of "economics" and "library and information sciences" to represent a second group of disciplines potentially less affected by the Chinese political system. Coverage of journals in the three subject categories in the Chinese domestic *CSSCI,* and the international *SSCI* is shown in Table 1.

Regarding the methodology, Pajek—available at http://vlado.fmf.uni-lj.si/pub/networks/pajek—is used for visualizing the citation networks. The cosine measure between two vectors is used as the similarity measure between the distributions for the various journals included in a citation environment. Cosine values below 0.2 are suppressed in the figures in order to enhance the interpretability of the visualizations. Because the cosine has values only between zero and one, low values of the cosine may indicate negative correlations (Egghe & Leydesdorff, 2009).

Within Pajek, the user can choose a variable width for the lines in the network, and colours or grey shades for nodes and links, respectively. The nodes can also be partitioned (and coloured) in accordance with their allocation into clusters using the various graph-analytical tools available within the program. In this study, we use the *k*-core algorithm as a first approximation for this delineation. The vertical size of the nodes is proportionate to the logarithm of the citations incurring to each node in the respective environments; horizontal axes of the nodes are diminished with self-citations. The width of the lines is proportionate to the cosine value of the association.

**Results**

a) The CSSCI and the SSCI

There remains a considerable difference between the *CSSCI* and the *SSCI,* despite the efforts of the *CSSCI* team in mirroring the structure of the *SSCI*. For example, the *CSSCI* is organized into only 25 subject categories, while the *SSCI* contains 54 such categories. Table 2 lists some descriptive statistics for both the *CSSCI* and the *SSCI*. In terms of the size of the two databases, the *SSCI* is obviously larger than the *CSSCI*. The number of journals in the *SSCI* is approximately four times that of the *CSSCI*.

**Table 2.** Comparison of the data in various dimensions for the *CSSCI* and the *SSCI* in 2007.

|  |  | *CSSCI* | *SSCI* | *SSCI/CSSCI* |
|---|---|---|---|---|
| *Number of source journals processed* | | 493 | 1,865 | 4 |
| *Citing* | Total number of citations | 227,456 | 3,672,282 | 16 |
| | Nr of citations per journal | 461 | 1969 | 4 |
| *Cited* | Total number of citations | 127,233 | 1,986,996 | 16 |
| | Nr of citations per journal | 258 | 1,065 | 4 |

As noted, Chinese scholars in the social sciences are less active in providing references (as in the natural sciences). In 2007, both the average citing counts per journal, and average cited counts per journal in the *CSSCI* are approximately 75% lower than in the *SSCI*. The two domains of cited and citing are "closed" sets because we did not include the non-source citations (which are available in both databases, but extremely sparse in the *CSSCI*).

b) *Citation patterns of Chinese and international journals in the social sciences*

b1. *Citation patterns of journals in political science*

The *Journal of Political Science* (政治学研究) and the journal *Marxism and Reality* (马克思主义与现实) in the *CSSCI,* are selected as seed journals. We use the former to represent Chinese journals in political science, while the latter can be considered typical for journals in marxism. In the *CSSCI*, marxism is an independent subject category, which means that marxism is considered as a field different from the political sciences. Journals focusing on political economy and philosophy are also subsumed under this category. With the special position of marxism in political science in China, it is worthwhile to investigate if journals in marxism and political science vary in terms of communication structures.

The journal *Political Analysis* (included in the international *SSCI*), was selected for a comparative study with the Chinese *Journal of Political Science* (政治学研究). There is no journal with the word "marxism" in its title in the *SSCI*, but there are two journals whose titles can be relevant to marxism (as different from western neo-marxism as a branch of philosophy): these journals are *Communist and Post-Communist Studies,* and *Problems of Post-Communism*. The former is selected for the comparison in this study since it has a higher impact factor and its focus is intellectually closer to that of the Chinese journal *Marxism and Reality*.

*Citing patterns of journals in political science*

The cluster of journals subsumed under marxism dominates the citing environment of *Marxism & Reality*. However, except for the seed journal itself, and the journal *Marxism Studies*, not one of the journals in the remainder of the cluster contains the word "marxism" in its title. The other journals can be classified into two types in terms of field focus: multidisciplinary and philosophy. This means that marxism studies in China are either multidisciplinary or dispersedly published in multidisciplinary journals. Marxism has a close relation to philosophy, and is relevant to fields in other social studies (Figure 2a).

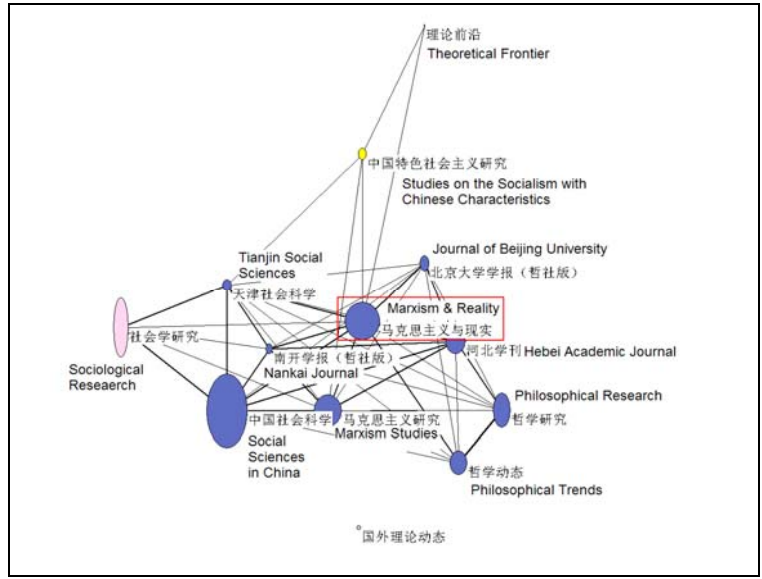

**Figure 2a.** Citing patterns of *Marxism & Reality* (马克思主义与现实) in 2007, threshold = 1%; cosine ≥ 0.2.

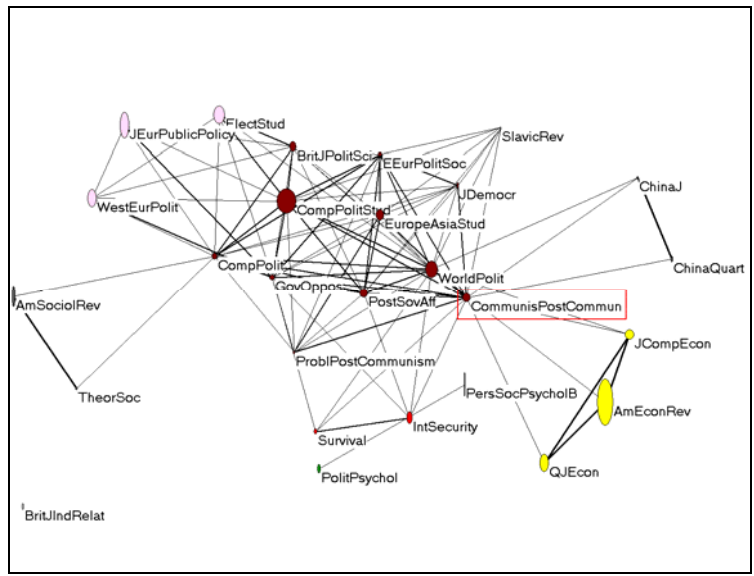

**Figure 2b.** Citing environment of *Communist and Post-Communist Studies* in 2007, threshold = 0%; cosine ≥ 0.2.

Compared to its Chinese counterpart, which provides references to many journals in the domestic arena, the international journal *Communist and Post-Communist Studies* contains fewer references to other journals in the *SSCI* database. Only two journals take 1% or more of the total references of *Communist and Post-Communist Studies*. This suggests—not surprisingly— that more journal resources with significant impact on marxism/communism studies are available within China, than in the international community. Most journal sources in the international community are from the political sciences instead of marxism studies, when all the referred journals are covered by setting the threshold at 0% (Figure 2b). *Communist and Post-Communist Studies,* and the other journals focusing on post-communism, mainly publish papers relevant to Eastern Europe (Soviet) and Asian (mainly the Chinese) systems.

Let us now compare two journals more central to political science in terms of their citing patterns. The citing environment of the *Journal of Political Science* (政治学研究) contains two clusters in the *CSSCI*. The cluster with the seed journal consists of six journals. Except for the seed journal, with a title focusing completely on research in political science, the other five journals are multidisciplinary and entertain different field interests according to their titles. Nevertheless, the six journals can be expected to maintain a common interest specifically in political science, which makes them relating in the same cluster in terms of their reference patterns (Figure 3a).

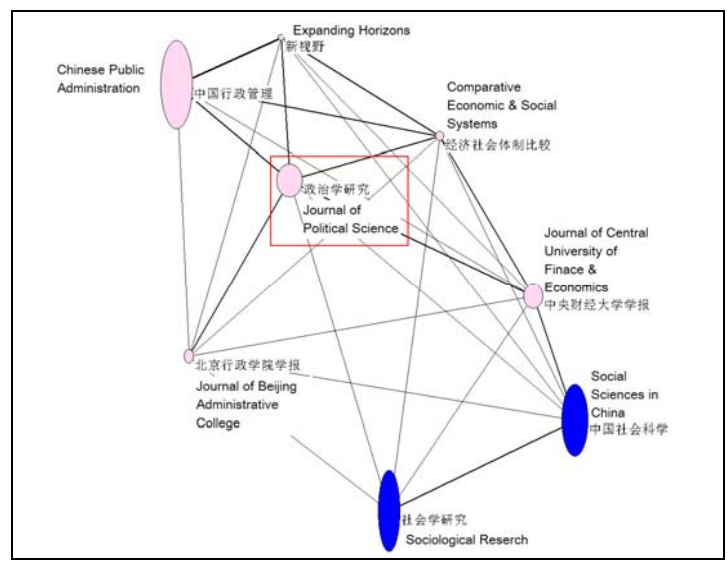

**Figure 3a.** Citing environment of the *Journal of Political Science* (政治学研究) in 2007, threshold =1%; cosine ≥ 0.2.

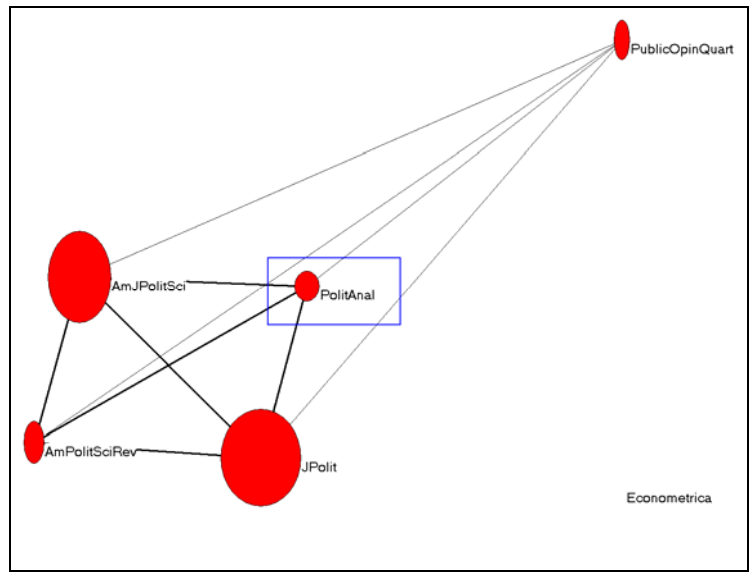

**Figure 3b.** Citing environment of *Political Analysis* in 2007, threshold = 1%; cosine ≥ 0.2.

Like *Marxism & Reality,* the citing environment of the domestic *Journal of Political Science* involves more than a single cluster with different publishing foci. However, the citing environment of the international journal, *Political Analysis,* is dominated by a single cluster with journals exclusively publishing research output in political science (Figure 3b).

In summary, few journals in China focus only on political science per se, which results in research outputs scattered in multidisciplinary journals. We found that political science in China is a multi-disciplinary field of study involving economic and social systems, economics, and public administration.

*Cited patterns of journals in political science*

The citation impact of *Marxism & Reality* in the Chinese domestic arena is multidisciplinary. Journals citing *Marxism & Reality* have similar intellectual interests (i.e., marxism), and are from different intellectual fields such as marxism studies, philosophy, public administration, and sociology. The Chinese journal, *Marxism & Reality,* also has impact on journals in law, dialectics of nature, morality and civilization, as well as economic issues. Compared to its citing environment, *Marxism & Reality* has a large impact at the field level (Figure 4a).

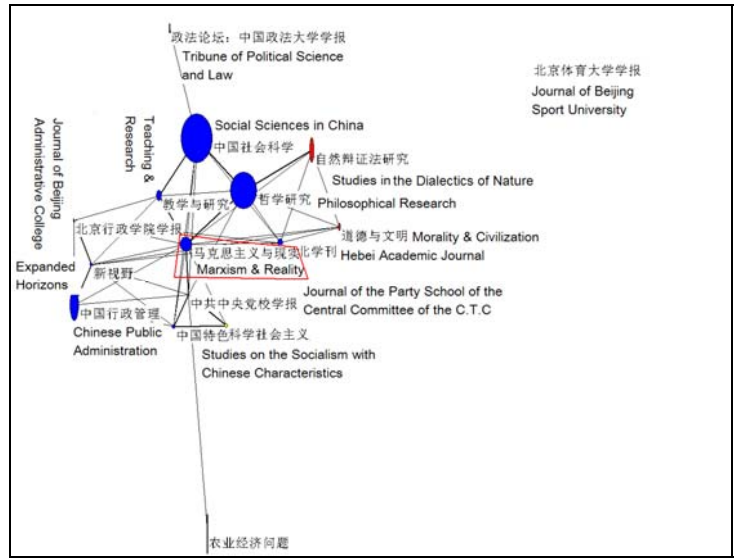

**Figure 4a.** Cited environment *Marxism & Reality* (马克思主义与现实) in 2007, threshold = 1%; cosine ≥ 0.2.

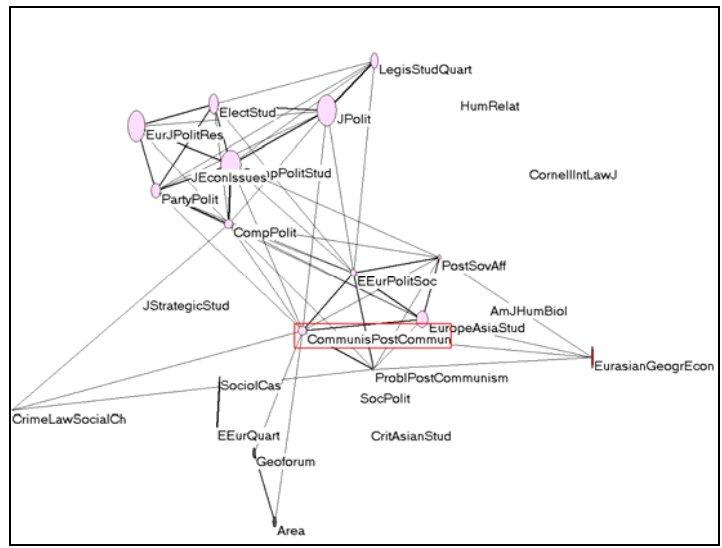

**Figure 4b.** Cited environment of the *Communist and Post-Communist* in 2007, threshold = 1%; cosine ≥ 0.2.

Although journal sources of communism studies were scarce in the international community, the citation impact of *Communist and Post-Communist Studies* is widespread: the journal is cited by journals in a variety of fields including political science, communist studies, law, economic issues, and other fields in the social sciences (Figure 4b). The seed journal, and journals close to it, have a comparable focus: they publish papers relevant to Eastern Europe (Soviet) and Asian (post-)communist systems.

The cited environment of the *Journal of Political Research* in the domestic arena is dispersed. Excepting the seed journal, other journals in the same cluster focus on public administration or economic and social issues. Journals in the other clusters have interests in studies in management, sociology, and so on. The citation impact of the seed journal is multidisciplinary, extending to several fields and forming several distinct clusters. In addition to being closely related to public administration, Chinese political science has impact on other fields in the social sciences and on management studies (Figure 5a).

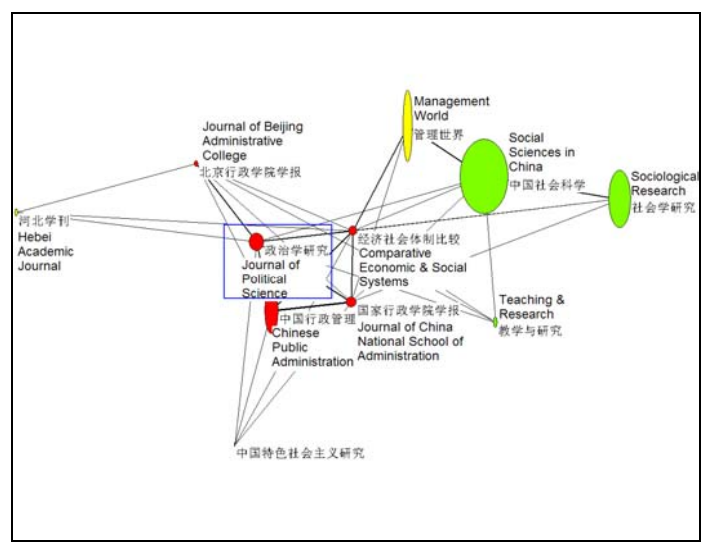

**Figure 5a.** Cited environment of the *Journal of Political Research* (政治学研究) in 2007, threshold = 1%; cosine ≥ 0.2.

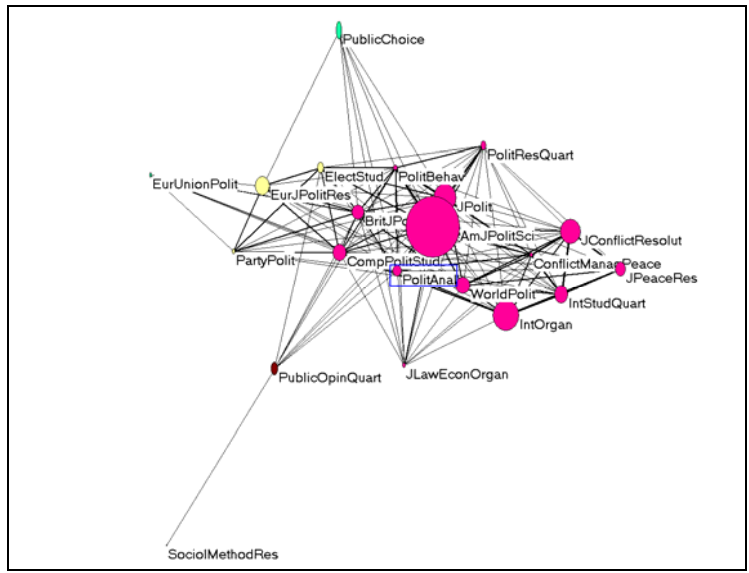

**Figure 5b.** Cited environment of the *Political Analysis* in 2007, threshold = 1%; cosine ≥ 0.2.

The cited environment of *Political Analysis* in the international community is more focused than its Chinese counterpart: the majority of these journals forms a core cluster with a focus on political science itself. International journals in political science also have an impact on public administration, but this influence is secondary compared to the impact on its own field, that is, political science (Figure 5b).

In summary, marxism or socialism studies in China should not be considered as only a field of political science because it entertains extensions to philosophy, public administration, and sociology. Studies in law, dialectics of nature, morality and civilization, as well as economic issues may also have links with marxism studies. Studies in the political sciences are not actively published in China yet. On the one hand, journals having citation relations with the seed journal in the political sciences are from different or multidisciplinary fields. This implies that there are not enough journal media solely focusing on political science so that scholars in this field have to publish in other fields. On the other hand, one could also hypothesize that insufficient research output in the political sciences results hitherto in limited journal space (Price, [1963] 1986; Beaver & Rosen, 1979b). It seems that the relation between marxism studies and political science is not as strong as the relation between marxism studies and philosophy.

Marxism study is more established than political science in China. Clusters in the citing/cited environments of the two Chinese journals, *Marxism & Reality* and *Political Studies,* indicate that the former's citation environments are more specialized than those of the latter. The citation environments of marxism studies show a dominant cluster which represents a specialty. In China, political science is more like a multidisciplinary field which involves various other fields. In other words, political science in China is not as specialized as in the international community.

In the international community, only a handful of journals focus on the study of communism. Because of the special position of marxism in China, journal sources for communism studies are more available in China, than in the international community. The citing environments of *Marxism & Reality* and the *Communist and Post-Communist Studie*s are very different. Most international journal sources of communism studies are from the political sciences rather than the specialty itself.

*b2. Citation patterns of journals in library and information science.*

The *Journal of the China Society for Scientific and Technical Information* (情报学报, *JCSSTI*) and the *Journal of Academic Libraries* (大学图书馆学报, *JAL*) were selected from the *CSSCI* as seed journals for the analysis in library and information science. Established under the aegis of the Chinese Society for Scientific and Technical Information, the *Journal of the China Society for Scientific and Technical Information* publishes a wide range of research output in information science and technology. Both journals have high citation impact in the subject category of library and information science. Corresponding to the two Chinese journals, the *Journal of the American Society for Information Science and Technology* (*JASIST*) and *Portal – Libraries and the Academy* from the social science version of the *JCR* were selected for comparison.

*Citing patterns of journals in information science*

In their citing environments, both the *JCSSTI* and the *JASIST* locate in a single dominant cluster, but differ in terms of field focus. The cluster of the *JCSSTI* is composed of journals in both library and information science, but that of the *JASIST* contains (given the threshold of 1%) exclusively journals in information science and technology. Furthermore, articles in the *JCSSTI* also refer to journals in research management and science studies (Figure 6a).

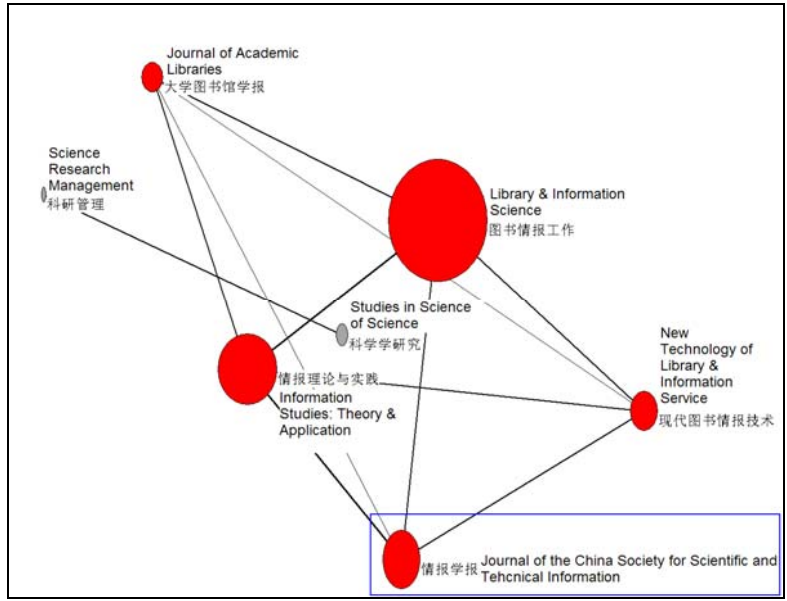

**Figure 6a.** Citing environment of the *Journal of the China Society for Scientific and Technical Information* (情报学报) in 2007, threshold = 1%; cosine ≥ 0.2.

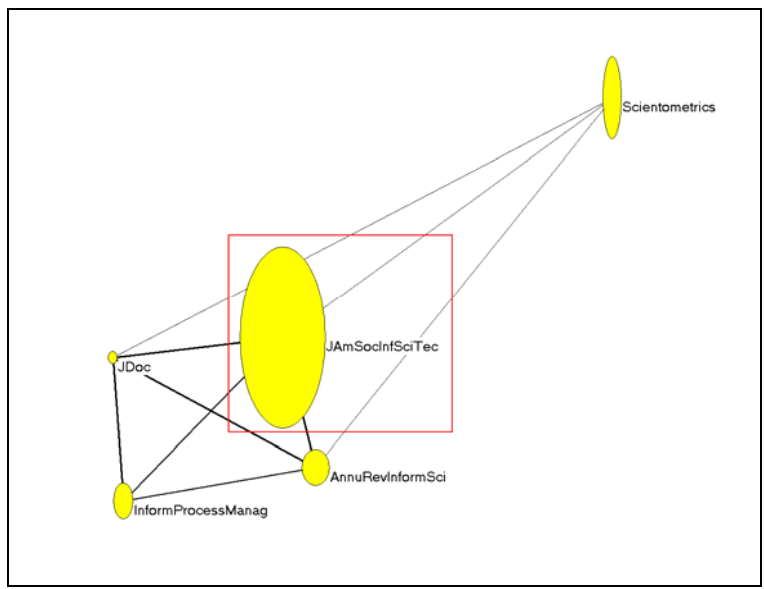

**Figure 6b.** Citing environment of the *Journal of the American Society for Information Science and Technology* in 2007, threshold = 1%; cosine ≥ 0.2.

Compared to the international community, Chinese research in information science seems to have closer relations with library science in terms of citing patterns, and is also more oriented toward research management and science studies than their international counterparts.

*Cited patterns of journals in information science*

The cited environment of the Chinese *JCSSTI* contains only a single cluster, which indicates that the journals can be considered as belonging to the same intellectual field. This equally applies to journals citing *JASIST*. However, journals citing the Chinese journal (*JCSSTI*) can be classified into both subfields: library and information science, while the citation impact of the international journal (i.e., the *JASIST*) is more exclusively within the information sciences (Figures 7a and b).

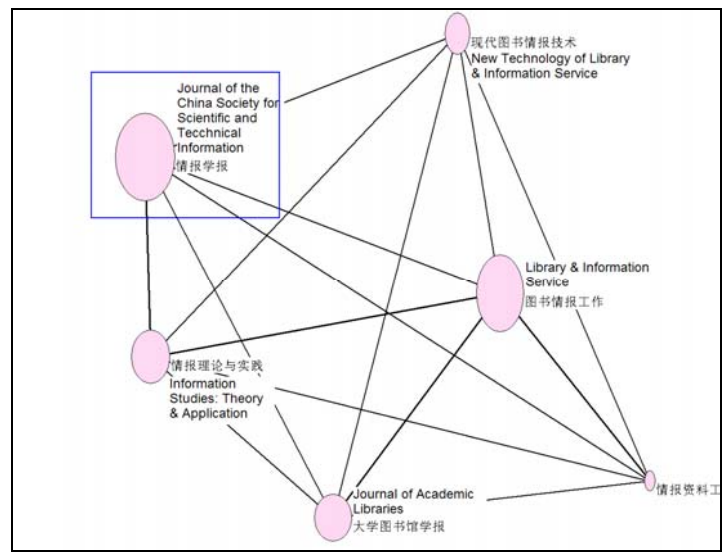

**Figure 7a.** Cited environment of the *Journal of the China Society for Scientific and Technical Information* (情报学报) in 2007, threshold = 1 %; cosine ≥ 0.2.

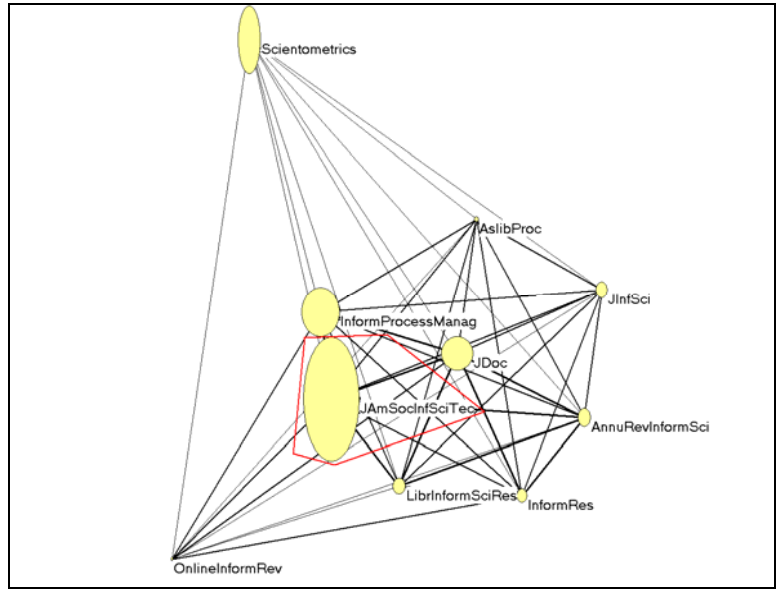

**Figure 7b.** Cited environment of the *Journal of American Society for Information Science and Technology* in 2007, threshold = 1%; cosine ≥ 0.2.

The citing and cited environments of Chinese and international journals in information science show that studies in library and information science are so closely dependent that they merge into one cluster or—in other words—form a single specialty in China, while in international studies the two fields are related, but distinguishable, from each other.

*Citation patterns of journals in library science*

Similar to the citing pattern of *JCSSTI,* which involves journals from both library and information science, the *Journal of Academic Libraries* also refers to journals from these two subfields. It becomes clear (from the figures not shown here) that journals in library and information science cite each other regularly in China. However, the international journal *Portal – Libraries and the Academy* mainly refers to journals in library science although two journals in information science are also involved in its impact environment. Thus, articles in library science contain references to journals in information science, but somewhat less so in the international arena than in the Chinese one.

The cited environment of Chinese journals in library science contains only a single cluster: journals in this environment have the same field focus. Within this cluster, library and information science penetrate into each other's domains. Most journals in either the citing/cited environments of the *Journal of the China Society for Scientific and Technical Information,* or the *Journal of Academic Libraries* are the same, which implies that these few journals play dominant roles in scholarly communication in both library and information science in China.

At the international level, although journals in library science are not so much cited by articles in *JASIST*, the impact of *JASIST* on library science is obvious (Figure 7b). Journals in library science do have impacts on journals in information science, and vice versa. However, journals in library science did not appear in the citing environment of the *JASIST* (Figure 6b) when the threshold was set at 1%. When the threshold is lowered to 0.5%, some journals in library science are present in the citing environment of *JASIST* as well.

In summary, Chinese and international journals in library and information science share similar citation patterns. Citation relations mainly happen within and across the two subfields. However, the citation relationship between the two subfields is denser in China than in the international community. Furthermore, Chinese journals in information science have visible involvement in research management and science studies. At the international level, these applications are organized in separate fields (e.g., business information systems).

*b3. Citation patterns of journals in economics*

The *Economic Research Journal* (*经济研究*) in the *CSSCI* and the *Quarterly Journal of Economics* in the *SSCI* were selected for investigating citation patterns of journals in the Chinese and international communities. Both journals have a high impact factor in the corresponding databases.

*Citing patterns of journals in economics*

In China, journals in economics form a dominant cluster in citing environments of Chinese journals. Journals in finance, management, and multidisciplinary research in the social sciences are also cited by articles published in economics journals. In other words, China's research in economics is not restricted to economics, but neighbouring fields like management, finance, accounting, and social issues are also cited (Figure 8a). The citing environment of international journals is more monodisciplinary, with journals in the same field, that is, economics, in the core set (Figure 8b).

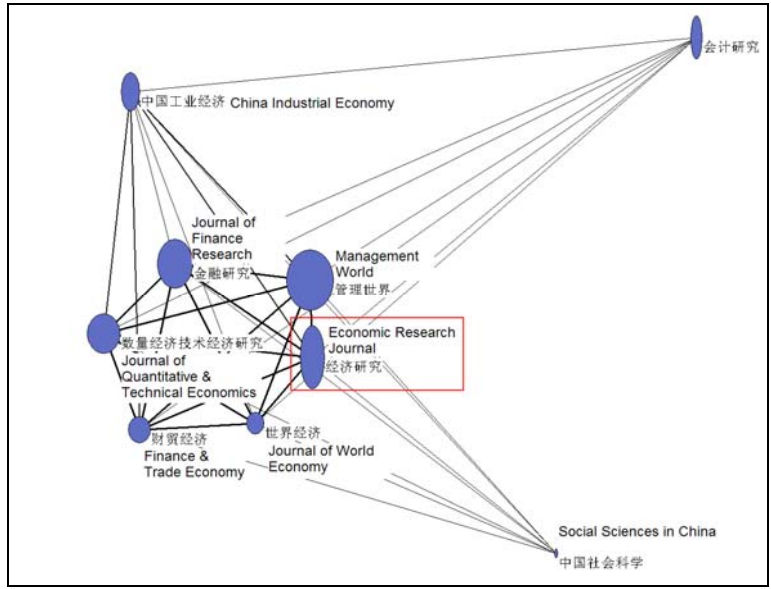

**Figure 8a.** Citing environment of the *Economic Research Journal* (经济研究) in 2007; cosine ≥ 0.2.

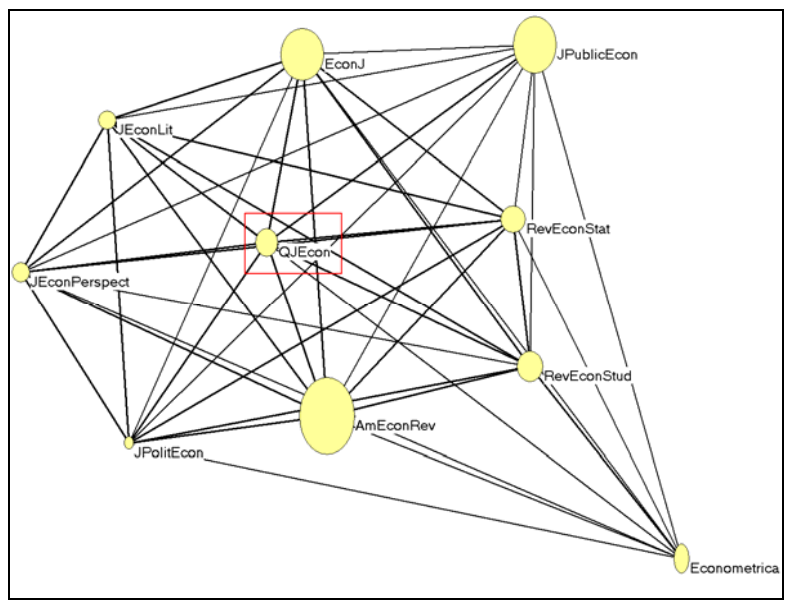

**Figure 8b.** Citing environment of the *Quarterly Journal of Economics* in 2007, threshold = 1%; cosine ≥ 0.2.

*Cited patterns of journals in economics*

In China, the impact of economic studies is multidisciplinary. The cited environment of the *Economic Research Journal* involves journals in economics, management, finance, accounting, and statistics (Figure 9a). At the international level, however, the impact of journals in economics is mainly in the economics field itself. Studies in finance may also refer to literature in economics (Figure 9b).

**Figure 9a.** Cited environment of the *Economic Research Journal* (经济研究) in 2007, threshold = 1%; cosine ≥ 0.2.

**Figure 9b.** Cited environment of the *Quarterly Journal of Economics* in 2007, threshold = 1%; cosine ≥ 0.2.

Chinese journals in economics are multidisciplinary, involving fields such as management, finance, accounting, and statistics, in addition to focusing on economics.

International journals in this field are more specialized with a core focus on economics. Studies in finance may also refer to literature in economics.

**Conclusions**

In this paper, we have investigated citation patterns of journals in three subject categories: political science including marxism/communism study, library and information science, and economics. We found that Chinese journals in the social sciences are organized differently from those in the international community. In summary, the Chinese journal structure in the social science does not yet exhibit specialization.

Studies in the social sciences have their own characteristics in China. Fields closely related to the Chinese political system are least specialized. Among the three disciplines under investigation, political science was least specialized, followed by economics. Library and information science, which has been least affected by the Chinese political system, is currently the most specialized among the three disciplines explored in our analyses. However, it seems that even disciplines without a clear relation to the political system are less specialized in China, than in the international community. These results suggest that Chinese studies in the social sciences are specialized less than their international counterparts.

Specialization can be considered as an important indicator of measuring the extent of knowledge accumulation in a discipline/field (Price,1963, 1986; Beaver & Rosen, 1979a, 1979b). A lower degree of specialization in the social sciences in China may be one of the reasons why China lags behind the West in this intellectual domain. China has already obtained second place in world scientific publications since 2006 (Zhou & Leydesdorff, 2008), but its international visibility in the social sciences has remained low (Zhou *et al*., 2009). The asynchronized internationalization between the natural and social sciences in China is astonishing.

**Discussion and some policy recommendations**

The lower degree of specialization of the social sciences in China can be attributed to a number of factors. First, the social sciences may have been more resistant to reform because of their deeper entrenchments in Chinese culture than the natural and life sciences. Since the introduction of the social sciences to China, at the time of the New Cultural Movement of 1915 (Cheng, 2004), cultural and ideological conflicts between the imported Western theories and the Chinese ideology have never ceased, resulting in an unsynchronized development of the social sciences between China and the West. However, such conflict may not exist for Chinese scholars who have years of Western education in their background.

Political movements, one after another, slowed down progress in the social sciences in China. After the establishment of the People's Republic of China in 1949, the social sciences experienced a flourishing period. Higher-education and research institutions were established successively. Policies encouraging these studies were implemented. However, during the Cultural Revolution, normal research in the social sciences came to a halt. It was not until 1978, when the debate on "practice is the sole criterion for testing truth" was launched, that institutions were reconstructed, and research and education in the social sciences were gradually recovered.

Nevertheless, in comparison to the natural sciences, the attention to reform in the social sciences has lagged behind, and thus the percentage of world share of publications has remained low. Government programs have not sufficiently paid attention to the social sciences. Whereas the Chinese government has two agencies at the national level engaged in the management and sponsoring of science and technology—the Ministry of Science and Technology (MOST), and the National Natural Science Foundation of China (NSFC)—the highest level of administration for the social sciences has been delegated to the National Planning Office of Philosophy and Social Sciences.

The responsible government agencies have not been active in developing a research evaluation system in the social sciences, as those have in the natural sciences. Without such an evaluation system, however, career opportunities may be determined by old boys networks and informal relations—"Quanxi" in Chinese (Zhu & Zhang, 2009). Academic corruption in Chinese scholarly communication has been frequently reported in recent years, due to the lack of an effective evaluation system (Cao, 2006; Wei, 2009; Xu, 2009; Wu, 2009). In our opinion, the establishment of a more objectifying system of research evaluation is urgent in the social sciences in order to improve research quality.

The internationalization of the domestic journals could be a second policy objective. More urgently than in the natural sciences, journals should be encouraged (if necessary with subsidies) to open their editorial policies to international participation; for example, by changing to a bilingual publication policy. A recent event shows China's positive attitudes to internationalization: a core journal of the Chinese Communist Party—求是 (i.e., *Seeking for the Truth*)—has begun to publish an English version (*Qiu Shi*) since October 2009, with the purpose of enabling the world to understand the values, ideology, theories and thoughts of the Chinese Communist Party (Qiu Shi Theory Net, 2009).

It is understandable that such reform policies will meet with resistance among Chinese intellectuals. The current state of the social sciences is comparable to that in the smaller countries of Europe during the 1970s. Reorganizations during the 1980s transformed, for example, the Dutch system from a locally oriented one to an international one (Van der Meulen & Leydesdorff, 1991). National journals are nowadays also specialized. Scandinavian countries went through similar transitions during this period, and Spain and Italy followed during the 1990s. French and German journals have become more specialized and are now sometimes multilingual.

Compared to the life and natural sciences, the social sciences are sometimes more local or national (Hicks, 1999, 2004; Kyvik, 1998). Furthermore, communication structures may vary extensively, especially between fields strongly localized or nationalized and fields more internationalized. The current study covers fields of both types and may therefore reflect the overall situation in China. Our results explain China's relative stagnation in emerging as a leading nation in the social sciences. The further synchronization between the Chinese domestic and international communication structures is important in a globalizing and increasingly knowledge-based economy.